\documentclass[published]{nst}
\usepackage{subfigure,dcolumn}
\usepackage[T2A,T1]{fontenc}
\usepackage[russian,english]{babel}

\usepackage{listings}
\usepackage{epstopdf}
\begin{document}

\title{Prediction for the synthesis cross sections of new moscovium isotopes in fusion-evaporation reactions}\thanks{Supported by National Natural Science Foundation of China (No. 12105241, 12175072), Natural Science Foundation of Jiangsu Province (No. BK20210788), Jiangsu Provincial Double-Innovation Doctor Program (No. JSSCBS20211013) and University Science Research Project of Jiangsu Province (No. 21KJB140026) and Lv Yang Jin Feng (No. YZLYJFJH2021YXBS130) and the Key Laboratory of High Precision Nuclear Spectroscopy, Institute of Modern Physics, Chinese Academy of Sciences (No. IMPKFKT2021001)}

\author{Peng-Hui Chen}
\email[Corresponding author,]{chenpenghui@yzu.edu.cn}
\affiliation{School of Physical Science and Technology, Yangzhou University, Yangzhou 225009, China}
\affiliation{Institute of Modern Physics, Chinese Academy of Sciences, Lanzhou 730000, China}

\author{Hao Wu}
\affiliation{School of Physical Science and Technology, Yangzhou University, Yangzhou 225009, China}


\author{Zu-Xing Yang}
\affiliation{RIKEN Nishina Center, Wako, Saitama 351-0198, Japan}


\author{Xiang-Hua Zeng}
\affiliation{School of Physical Science and Technology, Yangzhou University, Yangzhou 225009, China}
\affiliation{College of Electrical, Power and Energy Engineering, Yangzhou University, Yangzhou 225009, China }

\author{Zhao-Qing Feng}
\email[Corresponding author,]{fengzhq@scut.edu.cn}
\affiliation{School of Physics and Optoelectronics, South China University of Technology, Guangzhou 510641, China}

\begin{abstract}
In the framework of the dinuclear system model, the synthesis mechanism of the superheavy nuclides with atomic number $Z=112, 114, 115$ in the reactions of projectiles $^{40,48}$Ca bombarding on targets $^{238}$U, $^{242}$Pu, and $^{243}$Am at a wide incident energies (excitation energy from 0-100 MeV) have been investigated systematically. Based on the available experimental excitation functions, the dependence of calculated synthesis cross sections on collision orientations has been studied thoroughly. The TKEs of these collisions with the fixed collision orientation show its orientation dependence which can be used to predict the tendency of kinetic energy diffusion. The TKEs are dependent on incident energies which have been discussed. The method of Coulomb barrier distribution function has been applied in our calculations which could treat all of the collision orientations from the tip-tip to side-side approximately. The calculations of excitation functions  of $^{48}$Ca + $^{238}$U, $^{48}$Ca + $^{242}$Pu, and $^{48}$Ca + $^{243}$Am have a nice agreement with the available experimental data. 
The isospin effect of projectiles on production cross sections of moscovium isotopes and the influence of entrance channel effect on the synthesis cross sections of superheavy nuclei have been discussed.
The synthesis cross section of new moscovium isotopes $^{278-286}$Mc have been predicted as large as hundreds pb, in the fusion-evaporation reactions of $^{35,37}$Cl + $^{248}$Cf, $^{38,40}$Ar + $^{247}$Bk, $^{39,41}$K + $^{247}$Cm, $^{40,42,44,46}$Ca + $^{238}$Am, $^{45}$Sc + $^{242}$Pu, and $^{46,48,50}$Ti + $^{243}$Np, $^{51}$V + $^{238}$U at the excitation energy interval of 0-100 MeV.
\end{abstract}


\maketitle

\section{Introduction}\label{sec1}

Since the "island of stability" of superheavy nuclei has been predicted by the shell model in the 1960s\cite{Oganessian_2015}, the synthesis of superheavy nuclei were an exciting frontier field in the laboratories, which could provide a unique tool to explore the properties of nuclei and atomic structure under the extreme strong Coulomb force. However, due to the extremely low production cross sections, the synthesis of superheavy nuclei in current experiments spend long-time and cost lots of money. So it is particularly necessary to make reliable theoretical calculations that provide a reasonable reference for experiments. In recent years, to synthesize superheavy elements in the low energy heavy-ion collisions near Coulomb barrier has attracted extensive attentions from theorists and experimentalists.

On the experimental side, in past three decades, there were fifteen superheavy elements from Z = 104-118 have been synthesized and identified in the laboratories all over the world\cite{THOENNESSEN2013312}. Generally, the superheavy synthesis methods were classified by the excitation energy of compound nuclei as cold fusion and hot fusion, resulting in the compound nuclei survived by emitting 1-2 neutrons and 3-5 neutrons respectively, against fission. Based on cold fusion, in 1969, rutherfordium $^{257,258,259}$Rf was essentially discovered simultaneously in Dubna\cite{HEINLEIN1978407} and Berkeley\cite{GHIORSO197095} in the reactions of $^{249}$Cf($^{12,13}$C, 3-4n)$^{257,258,259}$Rf at the incident energy $E_{\rm lab}$ = 10.4 MeV/nucleon. Thirteen rutherfordium isotopes have been reported so far.
In 1970, $^{260,261}$Db was essentially discovered simultaneously in Dubna\cite{1970flog} and Berkeley\cite{PhysRevLett.24.1498} in the reactions of $^{249}$Cf($^{15}$N,4n)$^{260}$Ds at $E_{\rm lab}$ = 85 MeV and $^{243}$Am($^{22}$Ne, 4n)$^{261}$Ds at $E_{\rm lab}$ = 114 MeV. Up-to-now, eleven dubnium isotopes have been reported.
In 1974, $^{259,263}$Sg was essentially discovered simultaneously in Dubna\cite{1975Og} and Berkeley\cite{PhysRevLett.33.1490} in the reactions of $^{207}$Pb($^{54}$Cr,2n)$^{259}$Sg at $E_{\rm lab}$ = 262 MeV and $^{249}$Cf($^{18}$O, 4n)$^{263}$Sg at $E_{\rm lab}$ = 95 MeV. Twelve seaborgium isotopes have been discovered.
In 1981, $^{262}$Bh was essentially discovered in GSI\cite{1981ZPhyA300107M} in the reactions of $^{209}$Bi($^{54}$Cr, 1n)$^{262}$Bh at $E_{\rm lab}$ = 4.85 MeV/u. So far, ten bohriumium isotopes have been synthesized.
In 1984, $^{263-265}$Hs was essentially discovered simultaneously at GSI\cite{article84mu} in the reactions of $^{208}$Pb($^{58}$Fe, 2n)$^{265}$Hs at $E_{\rm lab}$ = 5.02 MeV/u. Hithergo, twelve hassium isotopes have been identified.
In 1982, $^{266}$Mt was essentially discovered simultaneously at GSI\cite{Mnzenberg1982} in the reactions of $^{209}$Bi($^{58}$Fe, 1n)$^{266}$Mt at $E_{\rm lab}$ = 5.15 MeV/u. Seven meitnerium isotopes have been obtained.
In 1995, $^{269}$Ds was essentially discovered simultaneously at GSI\cite{articlehof95} in the reactions of $^{208}$Pb($^{62}$Ni, 1n)$^{269}$Ds at $E_{\rm lab}$ = 311 MeV. Eight darmstadtium isotopes have been published up-to-now.
In 1995, $^{272}$Rg was essentially discovered simultaneously at GSI\cite{articlehof111} in the reactions of $^{209}$Bi($^{64}$Ni, 1n)$^{272}$Rg at $E_{\rm lab}$ = 318, 320 MeV. Seven roentgenium isotopes have been identified.
In 1996, $^{277}$Cn was essentially discovered simultaneously at GSI\cite{Hof96112} in the reactions of $^{208}$Pb($^{70}$Zn, 1n)$^{277}$Rg at $E_{\rm lab}$ = 344 MeV. Six copernium isotopes have been reported so far.
In 2004, $^{278}$Nh was essentially discovered simultaneously at RIKEN\cite{doi:10.1143/JPSJ.73.2593} in the reactions of $^{209}$Bi($^{70}$Zn, 1n)$^{278}$Nh at $E_{\rm lab}$ = 352.6 MeV. Six nihonium isotopes have been reported so far.
In 2004, $^{286-289}$Fl was essentially discovered simultaneously at Dubna\cite{PhysRevLett.105.182701} in the reactions of $^{244}$Pu($^{48}$Ca, 3-6n)$^{286-289}$Fl at $E_{\rm lab}$ = 352.6 MeV. Five flerovium isotopes have been reported so far.
In 2004, $^{288}$Mc was essentially discovered simultaneously at Dubna\cite{PhysRevC.69.021601} in the reactions of $^{243}$Am($^{48}$Ca, 3n)$^{288}$Mc at $E_{\rm lab}$ = 248, 253 MeV. Four moscovium isotopes have been reported.
In 2004, $^{286-289}$Lv was essentially discovered simultaneously at Dubna\cite{PhysRevC.69.054607} in the reactions of $^{245}$Cm($^{48}$Ca, xn)$^{293-x}$Fl at $E_{\rm lab}$ = 243 MeV. Four livermorium isotopes have been reported so far.
In 2011, $^{293-294}$Ts was essentially discovered simultaneously at Dubna\cite{PhysRevLett.104.142502} in the reactions of $^{249}$Bk($^{48}$Ca, 3-4n)$^{293-294}$Fl at $E_{\rm lab}$ = 247, 252 MeV. Two tennessine isotopes have been reported.
In 2006, $^{294}$Og was essentially discovered simultaneously at Dubna\cite{PhysRevC.74.044602} in the reactions of $^{249}$Cf($^{48}$Ca, 3n)$^{294}$Og at $E_{\rm lab}$ = 251 MeV. One oganesson isotope has been synthesized so far.
The synthesis information of the most neutron-rich and proton-rich superheavy nuceli with atomic number Z=104-118 were elements, isotopes, reactions, channels, laboratories and year, as illustrated in Table \ref{tab1}. Chinese superheavy group has synthesized the superheavy isotopes of $^{258,259}$Db\cite{epjagan01}, $^{264,265,266}$Bh\cite{06-4-8npr06} and $^{271}$Ds\cite{cpl12zhang} in IMP Lanzhou China. 

\begin{table}[!ht]
\renewcommand\arraystretch{1.5}
\centering
\footnotesize
\caption{\label{tab1} The synthesis of the most neutron-rich and proton-rich superheavy isotopes with atomic number Z=104-118 were illustrated as the production reactions, evaporation channel, synthesis laboratory, year and reference.}
\begin{tabular}{ccccccc}
\hline
Element & Isotopes & Reactions & Channel & Lab & Year & Ref.  \\ \hline
Rf(104) & $^{253}$ Rf & $^{50}$ Ti + $^{204}$ Pb & 1n & GSI & 1997 &\cite{Phys.A359(1997)415} \\
 & $^{267}$Rf & $^{48}$Ca + $^{242}$Pu & $\alpha$ & Dubna & 2004 &\cite{PhysRevC.69.064609} \\
 \hline
 Db(105) & $^{256}$Db & $^{50}$Ti + $^{209}$Bi & 3n & GSI & 2001 &\cite{Eur.Phys.J.A12(2001)57} \\
 & $^{270}$Db & $^{48}$Ca + $^{249}$Bk & 3n$\alpha$ & Berkely & 2010 &\cite{PhysRevLett.104.142502} \\
 \hline
 Sg(106)& $^{258}$Sg & $^{51}$V + $^{209}$Bi & 2n & GSI & 1997 &\cite{Phys.A359(1997)415} \\
 & $^{271}$Sg & $^{48}$Ca + $^{238}$U & $\alpha$ & Dubna & 2004 &\cite{PhysRevC.69.064609}\\
 \hline
 Bh(107)& $^{260}$Bh & $^{52}$Cr + $^{209}$Bi & $\alpha$ & Berkely & 2008 &\cite{PhysRevLett.100.022501} \\
 & $^{274}$Bh & $^{48}$Ca + $^{249}$Bk & 3n$\alpha$ & Dubna & 2010 &\cite{PhysRevLett.104.142502} \\
 \hline
Hs(108) & $^{263}$Hs & $^{56}$Fe + $^{208}$Pb & 1n & Berkely & 2009 &\cite{PhysRevC.79.011602} \\
 & $^{277}$Hs & $^{48}$Ca + $^{244}$Pu & 3n$\alpha$ & GSI & 2010 &\cite{PhysRevLett.104.252701} \\
 \hline
Mt(109) & $^{266}$Mt & $^{58}$Fe + $^{209}$Bi & 1n & GSI & 1982 &\cite{munzenberg1982observation} \\
 & $^{278}$Mt & $^{48}$Ti + $^{249}$Bk & 3n$\alpha$ & Dubna & 2010 &\cite{PhysRevLett.104.142502}\\
 \hline
 Ds(110)& $^{267}$Ds & $^{59}$Co + $^{209}$Bi & 1n & Berkely & 1995 &\cite{PhysRevC.51.R2293} \\
 & $^{281}$Ds & $^{48}$Ca + $^{244}$Pu & 3n$\alpha$ & Dubna & 2004 &\cite{PhysRevC.69.054607} \\
 \hline
Rg(111) & $^{272}$Rg & $^{64}$Ni + $^{209}$Bi & 1n & GSI & 1995 &\cite{Z.Phys.A350(1995)281} \\
 & $^{282}$Rg & $^{48}$Ca + $^{249}$Bk & 3n$\alpha$ & Dubna & 2010 &\cite{PhysRevLett.104.142502}\\
 \hline
Cn(112) & $^{277}$Cn & $^{70}$Zn + $^{208}$Pb & 1n & GSI & 1996 &\cite{Z.Phys.A354(1996)229} \\
 & $^{285}$Cn & $^{48}$Ca + $^{244}$Pu & 3n$\alpha$ & Dubna & 2004 &\cite{PhysRevC.69.054607}\\
 \hline
 Nh(113)& $^{278}$Nh & $^{70}$Zn + $^{209}$Bi & 1n & RIKEN & 2004 &\cite{Phys.Soc.Japan73(2004)2593} \\
 & $^{286}$Nh & $^{48}$Ca + $^{249}$Bk & 3n$\alpha$ & Dubna & 2010 &\cite{PhysRevLett.104.142502} \\
 \hline
 FI(114)& $^{285}$FI & $^{48}$Ca + $^{242}$Pu & 5n & Berkely & 2010 &\cite{PhysRevLett.105.182701} \\
 & $^{289}$FI & $^{48}$Ca + $^{244}$Pu & 3n & Dubna & 2004 &\cite{PhysRevC.69.054607} \\
 \hline
 Mc(115)& $^{287}$Mc & $^{48}$Ca + $^{243}$Am & 4n & Dubna & 2004 &\cite{PhysRevC.69.021601} \\
 & $^{290}$Mc& $^{48}$Ca + $^{249}$Bk & 3n$\alpha$ & Dubna & 2010 &\cite{PhysRevLett.104.142502}\\
 \hline
 Lv(116)& $^{290}$Lv & $^{48}$Ca + $^{245}$Cm & 3n& Dubna & 2004 &\cite{PhysRevC.69.054607} \\
 & $^{293}$Lv & $^{48}$Ca + $^{245}$Cm & 1n & Dubna & 2004 &\cite{PhysRevC.69.054607} \\
 \hline
Ts(117) & $^{293}$Ts & $^{48}$Ca + $^{249}$Bk & 4n & Dubna & 2010 &\cite{PhysRevLett.104.142502} \\
 & $^{294}$Ts & $^{48}$Ca + $^{249}$Bk & 3n & Dubna & 2010 &\cite{PhysRevLett.104.142502} \\
 \hline
 Og(118)& $^{294}$Og & $^{48}$Ca + $^{249}$Cf & 3n & Dubna & 2006 &\cite{PhysRevC.74.044602} \\
 \hline
\end{tabular}
\end{table}

The mechanism of fusion-evaporation could not reach the next new period in the periodic table of elements easily because of the limited available combinations of projectile-target. With the development of suitable separation and detection techniques, the multinucleon transfer (MNT) mechanism might be as the most promising method used to synthesize unknown superheavy elements, which has been applied to produce massive heavy and superheavy isotopes\cite{VOLKOV197893}. 
The laboratories all over the world such as Institute of Modern Physics (IMP)\cite{arti22gan}, Gesellschatt F\"{u}r Schwerionenforschung (GSI) \cite{epjaheinz22,2022NC279B} and DUBNA\cite{inproceedings,epja20ga}, RIKEN\cite{KAJI2008198,nc1919,PhysRevLett.124.052502}, the Lawrence Berkeley National Laboratory (LBNL)\cite{PhysRevLett.124.252502,PhysRevLett.124.252502}, Australian National University (ANU)\cite{BANERJEE2021136601} devoted constantly to synthesize new superheavy elements and their isotopes. From the chart of nuclei, in the superheavy region, there are substantial isotopes of superheavy elements which are still unknown yet. One aim of this paper is to predict the production cross sections of moscovium isotopes in fusion-evaporation reactions based on the different combinations of projectile-target.

In the theoretical side, to depict the production mechanism of superheavy nuclei, some theoretical models have been built, for example, time-dependent Hartree-Fock (TDHF) \cite{GUO2018401,10.3389/fphy.2019.00020,MARUHN20142195}, the improved quantum molecular dynamics (ImQMD) model\cite{PhysRevC.65.064608,PhysRevC.88.044611,PhysRevC.88.044605}, the dynamical approach based on Langevin equations\cite{PhysRevC.96.024618,PhysRevC.85.014608}, the dinuclear system model (DNS)\cite{FENG200650,PhysRevC.91.011603,PhysRevC.89.024615,epja20gga,07fengcpl} $etc.$. The calculations of these theoretical models basically could have a good agreement with the available experimental data in which have their own features. In this work, the DNS model have been applied, which have some featured advantages such as the better consideration of
shell effect, dynamical deformation, fission, quasi-fission, deep-inelastic and odd-even effect, and its calculation efficiency is very high. In previous work\cite{PhysRevC.91.011603,FENG200650,FENG201082,07fengcpl}, the DNS model have nicely reproduced the available experiment results and predicted the synthesis production cross sections of superheavy elements and exotic heavy nuclei in the mechanisms of fusion-evaporation and MNT reactions.

In this paper, we have investigated the dependence of the evaporation residue cross sections on collision orientations and the influence of entrance channels effect on the evaporation residue cross sections. We have proposed the Gaussian-like barrier distribution function for treating the problem of collision orientation dependence.   
The article is organized as follows: In Sec. \ref{sec2} we give a brief description of the DNS model. Calculated results and discussions are presented in Sec. \ref{sec3}. Summary is concluded in Sec. \ref{sec4}.

\section{Model Description}\label{sec2}

Initially, the dinuclear system (DNS) concept were proposed to depict the deep inelastic reaction mechanism which was a molecular-like configuration of two colliding partners, keeping their own individuality in the collision process.
The DNS model has been used to describe the fusion-evaporation reactions and multinucleon transfer reactions widely. 
The complete fusion evaporation reaction can be depicted as the three process. Firstly, the colliding partners overcome the Coulomb barrier to form the composite system. Secondly, the kinetic energy and angular momentum dissipating into the composite system to enable the nucleons transfer between the touching colliding partners. Finally, all the nucleons have been transferred from projectile nuclei to the target nuclei which could form the compound nuclei with a few excitation energy and angular momentum. The highly excited compound nuclei will be de-excited by evaporating the light particles ( i.e. neutrons, $\gamma$-rays and light charged particles) or fission. Based on the DNS model, the evaporation residual cross sections of superheavy nuclei can be written as
\begin{eqnarray}
\sigma _{\rm ER}\left ( E_{\rm c.m.} \right ) =\frac{\pi \hbar ^2}{2\mu E{\rm c.m.}} \sum_{J=0}^{J_{\rm max}}(2J+1)T(E_{\rm c.m.},J)\nonumber \\P_{\rm CN}(E_{\rm c.m.},J)W_{\rm sur}(E_{\rm c.m.},J) 
\end{eqnarray}
Where, the penetration probability $T(E_{\rm c.m.},J)$ is the probability of the collision system passing through the Coulomb barrier which was calculated by the empirical coupling channel model\cite{FENG200650}. The fusion probability $P_{\rm CN(E_{\rm c.m.},J)}$ is the probability to form compound nuclei\cite{PhysRevC.80.057601,PhysRevC.76.044606}. The survival probability $W_{\rm sur}$ is the probability of the highly excited compound nuclei surviving by evaporating light particles against fission. The maximal angular momentum was set as $J_{\rm max}$ = 30-50, because the fission barrier for the superheavy nuclei may vanish at the high spin \cite{J.Mod.Phys.E5191(1996)}.

\subsection{ Capture probability}

The capture cross sections of the two colliding partners was given as
\begin{eqnarray}
\sigma _{\rm cap}(E_{\rm c.m.})=\frac{\pi \hbar ^2}{2\mu E_{\rm c.m.}}\sum_{J}^{}(2J+1)T(E_{\rm c.m.},J). 
\end{eqnarray}
Here, the penetration probability $T(E_{\rm c.m.,J})$ is evaluated by the Hill-Wheeler formula \cite{PhysRev.89.1102} with the barrier distribution function.
\begin{eqnarray}
 && T(E_{\rm c.m.},J)= \nonumber \int   f(B) \\&& \frac{1}{1+\exp\left \{ -\frac{2\pi }{\hbar \omega (J)}\left [ E_{\rm c.m.}-B-\frac{\hbar^2J(J+1)}{2\mu R\rm_B^2(J) }  \right ]  \right \} }dB. 
\end{eqnarray}
Here $\hbar \omega (J)$ is the width of the parabolic barrier at the position $R_{\rm B}(J)$. The normalization constant is with respect to the relation $\int f(B)dB=1$. The barrier distribution function is assumed to be in an asymmetric Gaussian form\cite{FENG200650,PhysRevC.65.014607}
\begin{eqnarray}\label{gbd}
f(B)=
\left\{\begin{matrix}
\frac{1}{N}\exp [-(\frac{B-B_{\rm m}}{\Delta_{1} } )]\ \ B<B_{\rm m},\\
  & \\\frac{1}{N}\exp [-(\frac{B-B_{\rm m}}{\Delta_{2} } )]\ \ B>B_{\rm m},
  &
\end{matrix}\right.
\end{eqnarray}
Here $\bigtriangleup _2$ = (B$_{0}$-B$_{\rm s}$)/2, $\bigtriangleup _1 $=$\bigtriangleup _2$-2 MeV, B$_m$=(B$_0$+B$_{\rm s}$)/2, B$_0$ and B$_{\rm s}$ are the Coulomb barriers of the side-side collision and the saddle point barriers in dynamical deformations\cite{PhysRevC.65.014607}. The nucleus-nucleus interaction potential was calculated by
\begin{eqnarray}
V(\{\alpha\})=V{\rm _C}(\{\alpha\})+V_{\rm N}(\{\alpha\}) + V_{\rm def}
\end{eqnarray}
with 
\begin{eqnarray}
V_{\rm def} = \frac{1}{2} C_1(\beta_1-\beta^0_1)^2+\frac{1} {2}C_2(\beta_2-\beta_2^0)^2). \nonumber
\end{eqnarray}

The 1 and 2 represent the projectile and the target, respectively.  $R=R_1+R_2+s$ and s are the distance between the center and the surface of projectile-target. $R_1$ and $R_2$ are the radii of the projectile and target,  respectively. The $\beta_{1(2)}^0$ are the static deformation of projectile-target. The $\beta_{1(2)}$ are the adjustable quadrupole deformation in which it was varied to find the minimal $V(\{\alpha\})$. The $\{\alpha\}$ stand for $\{R,\beta_1,\beta_1,\beta_2,\theta_1,\theta_2\}$. To reduce the number of deformation variables, we assume that the deformation energy of colliding system were proportional to their mass\cite{PhysRevC.65.014607}, namely, $C_1\beta_1^2/C_2\beta_2^2=A_1/A_2$. So we could use only one deformation parameter as $\beta=\beta_1+\beta_2$. The stiffness parameters $C_{\rm i}(\rm i=1,2)$ were calculated by the liquid-drop model\cite{MYERS19661}, as parameterization formula
\begin{eqnarray}
C\rm _i=(\lambda -1)\left [ (\lambda -1)R_i^2\sigma -\frac{3}{2\pi } \frac{Z^2e^2}{R_i(2\lambda+1)}  \right ] 
\end{eqnarray}
Here, the $ R\rm_{i}$ is the radius of the spheroidal nucleus which has the formula $R\rm_{i}$=1.18$A\rm_i^{1/3}$ (i=1,2)
In this work, the quadrupole deformation was taken into account $(\lambda=2)$. 
The $\sigma$ is the coefficient of surface tension which fits $4\pi_{\rm i}^2$$\sigma$=$a_{\rm s}A_{\rm i}^{2/3}$ where the $a_{\rm s}$ = 18.32 MeV is surface energy. The nuclear potential is calculated by the double-folding method \cite{PhysRevC.80.057601,PhysRevC.76.044606, J.Mod.Phys.E5191(1996)}.
\begin{eqnarray}\label{vn}
V_{\rm N}= && C_0\left \{ \frac{F\rm_{\rm in}-F\rm _{\rm ex}}{\rho _0}\left [ \int \rho _1^2(r)\rho_2(r-R)dr + \int \rho_1(r) \right.\right. \nonumber \\
 && \left. \left.  \rho_2^2(r-R)dR \right ]+F_{\rm ex}\int \rho _1(r)\rho _2(r-R)dr\right \} 
\end{eqnarray}
with 
\begin{eqnarray}
F\rm _{in(ex)}=f\rm_{in(ex)}+f_{in(ex)}^,\frac{N_1-Z_1}{A_1} \frac{N_2-Z_2}{A_2}. \nonumber
\end{eqnarray}
It depends on the nuclear density and orientation of the deformed colliding partners. The parameters $C_0 = 300$ MeV fm$^3$, $f_{\rm in}$ = 0.09, $f_{\rm ex}=-2.59$, $f_{\rm in}^{,}$ = 0.42, $f_{\rm ex}^{,}$ = 0.54, and $\rho_0$ = 0.16 fm$^3$ were used in our calculations. The Woods-Saxon density distribution was presented as:
\begin{eqnarray}
\rho _1(r)=\frac{\rho_0}{1+\rm exp\left [ (r-\Re _1(\theta_1)) \right/a_1 ] } 
\end{eqnarray}
and
\begin{eqnarray}
\rho _2(r-R)=\frac{\rho_0}{1+\rm exp\left [ (|r-R|-\Re _2(\theta _2))/a_2 \right ] } 
\end{eqnarray}
Here, $\Re_{\rm i}(\theta_{\rm i})(i=1,2)$ were the surface radii of the nuclei with the formula $\Re_{\rm i}(\theta_{\rm i})$=$\Re \left [ 1+\beta _{\rm i}Y_{20}(\theta _{\rm i}) \right ]$ where the $R_i$ was spheroidal nuclei radius. The $a_{\rm i}$ was the surface diffusion coefficient which was taken as 0.55 fm in our calculations. The Coulomb potential was derived by the Wong's formula, as follows \cite{PhysRevLett.31.766}.
\begin{eqnarray}\label{vc}
&&V_{\rm C}(\{\alpha\}) = \frac{Z_1Z_2e^2}{r} \nonumber \\
&&+\left ( \frac{9}{20\pi}  \right )^{1/2}\left ( \frac{Z_1Z_2e^2}{r^3} \right )
 \sum\rm_{i=1}^{2}R_i^2\beta\rm _iP_2(\cos\theta_i)
\nonumber \\
&& +\left( \frac{3}{7\pi } \right) \left( \frac{Z_1Z_2e^2}{r^3}\right) 
\sum\rm_{i=1}^{2}R_i^2 \left( \beta_iP_2 \cos\theta_i \right) ^2    
\end{eqnarray}

where $\theta_i$, $\beta_i$, $R_i$, and $P_2(\cos\theta_i)$ are the angle between the symmetry axis of the deformed projectile-target and collision axis, quadrupole deformation, the radius of the projectile-target, and Legendre polynomial, respectively. The Wong's formula was in good agreement with the double-folding method.

\subsection{Fusion probability}

The composite system has been formed after the capture process where the dissipation of kinetic energy and angular momentum happened, to activate nucleons transfer between the touching configuration of the projectile-target which result in the mass probability diffusion. 
The mass probability of the formed fragments was evaluated by solving a set of master equations. The term of mass probability $P(Z_1,N_1,E_1,t)$ contain that proton number, neutron number of $Z_1,$ and $N_1$, and the internal excitation energy of $E_1$ for given fragment $A_1$. The master equation was shown as \cite{PhysRevC.76.044606,FENG201082,FENG200933}
\begin{eqnarray}
&& \frac{d P(Z_1,N_1,E_1,t)}{d t} =  \nonumber \\ && \sum \limits\rm_{Z'_1}W\rm_{Z_1,N_1;Z'_1,N_1}(t) [d\rm_{Z_1,N_1}P(Z'_1,N_1,E'_1,t) \nonumber \\ && - d\rm_{Z'_1,N_1}P(Z_1,N_1,E_1,t)] + \nonumber \\ &&
 \sum \limits\rm_{N'_1}W_{Z_1,N_1;Z_1,N'_1}(t)[d\rm_{Z_1,N_1}P(Z_1,N'_1,E'_1,t) \nonumber \\ && - d\rm_{Z_1,N'_1}P(Z_1,N_1,E_1,t)] - \nonumber \\
 &&[\Lambda ^{qf}\rm_{A_1,E_1,t}(\Theta) + \Lambda^{fis}\rm_{A_1,E_1,t}(\Theta)]P(Z_1,N_1,E_1,t).
\end{eqnarray}
Here the $W_{Z_1,N_1,Z'_1,N_1}$ ($W_{Z_1,N'_1,Z_1,N_1}$) was the mean transition probability from the channel ($Z_1,N_1,E_1$) to ($Z'_1,N_1,E'_1$) [or ($Z_1,N_1,E_1$) to ($Z_1,N'_1,E'_1$)]. The $d_{Z_1,N_1}$ denotes the microscopic dimension corresponding to the macroscopic state ($Z_1,N_1,E_1$). The sum contains all possible numbers of proton and neutron for the fragment $Z'_1$, $N'_1$ own. However, only one nucleon transfer at one time was supposed in the model with the relation $Z'_1$ = $Z_1$ $\pm$ 1, and $N'_1$ = $N_1$ $\pm$ 1. The excitation energy $E_1$ was the local excitation energy $\varepsilon^*_1$ for the fragment ($Z'_1$, $N'_1$) which was derived by the dissipation of the relative motion along with PES of the DNS \cite{PhysRevC.27.590}. The time of dissipation process was evaluated by the parameterization classical deflection function \cite{LI1981107}. The motion of nucleons in interaction potential was governed by the single-particle Hamiltonian, 
\begin{eqnarray}
H(t) = H_0(t) + V(t)
\end{eqnarray}
where the total single particle energy and interaction potential were
\begin{eqnarray}
H_0(t) &&= \sum _K\sum_{\nu_K} \varepsilon_{\nu_K}(t)\alpha^+_{\nu_K}(t)\alpha_{\nu_K}(t) \\ 
 V(t) &&= \sum\rm _{K,K^{'}} \sum_{\alpha_K,\beta\rm_{K'}} u\rm_{\alpha_K,\beta_{K'}}\alpha^+_{\alpha\rm_K}(t)\alpha\rm_{\beta_K}(t) \nonumber  \\ 
 &&= \sum\rm_{K,K'}V_{K,K'}(t). 
\end{eqnarray}
 The quantities $\varepsilon_{\nu K}$ and $u_{\alpha_K,\beta_{K'}}$ represent the single particle energies and the interaction matrix elements, respectively in which the single-particle state was defined as the centers of colliding nuclei assumed to be orthogonal in the overlapping region. Then the annihilation and creation operators were dependent on time. The single-particle matrix elements were parameterized as
\begin{eqnarray}
&& u\rm_{\alpha_K,\beta_K'} =  U_{K,K'}(t)  \\ && \times \left\{ \rm exp \left[- \frac{1}{2}( \frac{\varepsilon\rm_{\alpha_K}(t) - \varepsilon\rm_{\beta_K}(t)}{\Delta\rm_{K,K'}(t)})^2 \right] - \delta\rm_{\alpha_K,\beta_{K'}} \right\} \nonumber
\end{eqnarray}
Here, The calculation of the $U_{\rm K,K'}(t)$ and $\delta_{\alpha_{\rm K},\beta_{\rm K'}}(t)$ have been described in Ref.\cite{PhysRevC.68.034601}.
The proton transition probability was microscopically derived by 
\begin{eqnarray}
\label{trw}
&& W_{Z_{1},N_{1};Z_{1}^{\prime},N_{1}} = \frac{\tau_{\rm mem}(Z_1,N_1,E_1;Z_1^{\prime},N_{1},E_{1}^{\prime})} {d_{Z_1,N_1} d_{Z_1 ^{\prime},N_1}\hbar^2} \nonumber \\
&&  \times \sum_{ii^{\prime}}|\langle  Z_{1}^{\prime},N_{1},E_{1}^{\prime},i^{\prime}|V|Z_{1},N_{1},E_{1},i \rangle|^{2}.
\end{eqnarray}
The neutron transition probability has the similar formula. The memory time and the interaction elements $V$ could be seen in the Ref.\cite{PhysRevC.80.057601}. 

The evolution of the DNS along the distance $R$ lead to quasi-fission. The decay probability of quasi-fission were calculated based on the one-dimensional Kramers equation as \cite{PhysRevC.68.034601,PhysRevC.27.2063}
\begin{eqnarray}\label{qf}
\Lambda^{qf}\rm_{A_1,E_1,t}(\Theta) = && \frac{\omega}{2\pi\omega^{B\rm_{qf}}}\left[\sqrt{(\frac{\Gamma}{2\hbar})^2 + (\omega^{B\rm_{qf}})^2} - \frac{\Gamma}{2\hbar}\right] \nonumber \\ && \times \rm exp\left[- \frac{B\rm_{qf}(A_1)}{\Theta(A_1,E_1,t)}\right]
\end{eqnarray}
where the $B_{\rm qf}(A_1)$ was the quasi-fission barrier. The $\omega$ and $\omega^{B_{\rm qf}}$ were the frequencies of the harmonic oscillator approximation at the bottom and top of the interaction potential pocket which were constants as $\hbar \omega^{B\rm_{qf}}$ = 2.0 MeV and $\hbar \omega$ = 3.0 MeV in this work. The $\Gamma$ = 2.8 MeV was the quantity characterizing the average double width of the single-particle state. The local temperature was given by the Fermi gas model $\Theta=(\varepsilon ^*/(A/12))^{1/2}$. In the nuclear collision process, heavy fragments might lead to fission where the fission probability were calculated by the Kramers formula 
\begin{eqnarray}
\Lambda^{\rm fis}\rm_{A_1,E_1,t}(\Theta) = && \frac{\omega_{\rm g.s}}{2\pi\omega\rm_f}\left[\sqrt{(\frac{\Gamma_0}{2\hbar})^2 + (\omega\rm_f)^2} - \frac{\Gamma_0}{2\hbar}\right] \nonumber \\ && \times \exp\left[- \frac{B\rm_f(A_1)}{\Theta(A_1,E_1,t)}\right]
\end{eqnarray}
where the $\omega_{\rm g.s.}$ and $\omega_f$ were the frequencies of the oscillators approximating the fission-path potential at the ground state and the top of the fission barrier for fragment $A_1$, respectively, which were set as $\hbar\omega_{\rm g.s.}$ = $\hbar\omega_{\rm f}$ = 1.0 MeV and $\Gamma_0$ = 2 MeV. The fission barrier was calculated by the macroscopic part plus the shell correction energy. 
In relaxation process of the relative motion, the DNS will be excited by the dissipation of the relative kinetic energy and angular momentum. The excited composite system opens a valence space $\Delta \varepsilon_K$ in fragment $K$ ($K$ = 1, 2) which has a symmetrical distribution around the Fermi surface. The nucleons in the valence space were actively enable to be excited and transfer. The averages on these quantities are performed in the valence space, 
\begin{eqnarray}
\Delta \varepsilon\rm_K = \sqrt{\frac{4\varepsilon^*\rm_K}{g\rm_K}},\quad
\varepsilon^*_K =\varepsilon^*\frac{A\rm_K}{A}, \quad
g_K = A_K /12,
\end{eqnarray}
where the $\varepsilon^*$ is the local excitation energy of the DNS, which provide the excitation energy
for the mean transition probability. There are $N_K$ = $g_{\rm K}\Delta\varepsilon_{\rm K}$ valence states and $m_{\rm K}$ = $N_{\rm K/2}$ valence nucleons in the valence space $\Delta\varepsilon_K$, which gives the dimension
\begin{eqnarray}
 d(m_1, m_2) = {N_1 \choose m_1} {N_2 \choose m_2}.
\end{eqnarray}
The local excitation energy is given by
\begin{eqnarray}
\varepsilon^* = E\rm_x - (U_{\rm dr}(A_1,A_2) - U_{\rm dr}(A_{\rm P}, A_{\rm T}))
\end{eqnarray}
Where the $U_{\rm dr}(A_1, A_2)$ and $U_{\rm dr}(A_P, A_T)$ were the driving potentials of fragments $A_1$, $A_2$ and $A_P$, $A_T$, respectively. The detailed calculation of the driving potentials is from  Eq. \ref{pes}. The excitation energy $E_{\rm x}$ of the composite system was converted from the relative kinetic energy dissipation\cite{PhysRevC.76.044606}. The potential energy surface (PES) of the DNS was written as
\begin{eqnarray}\label{pes}
&&U_{\rm dr}(A_1,A_2;J,\theta_1,\theta_2) =  B_1+B_2 - B\rm_{CN}-V^{\rm CN}_{\rm rot}(J) \nonumber\\&&
+V_{\rm C}(A_1,A_2;\theta_1,\theta_2) + V_{\rm N}(A_1,A_2;\theta_1,\theta_2)
\end{eqnarray}

 Here $B_{i}$ ($i$ = 1, 2) and $B_{\rm CN}$ were the negative binding energies of the fragment $A_{i}$ and the compound nucleus A = A$_1$+A$_2$, respectively, where the shell and the pairing corrections were included reasonably. The $V_{\rm rot}^{CN}$ is the rotation energy of the compound nuclei. The $\beta_{\rm i}$ represent the quadrupole deformations of binary fragments. The $\theta_{\rm i}$ denote collision orientations. The $V_{\rm C}$ and $V_{\rm N}$ were derived by Eq. \ref{vc} and Eq. \ref{vn}, respectively.

By solving a set of master equations, the probability of all possible formed fragments were presented. The hindrance in the fusion process named inner fusion barrier $ B_{\rm fus}$ which was defined by the difference from the injection position to the B.G. point. In the DNS model, these fragments overcoming the inner barrier which were considered to lead to fusion. Therefore, the fusion probability were evaluated by adding all of the fragments which could penetrate the inner fusion barrier. The fusion probability with the barrier distribution was evaluated by  
\begin{eqnarray}
P_{\rm CN}(E_{\rm c.m.},J,B)=\sum\rm_{A=A\rm_{\rm BG}}^{A\rm_{\rm CN}} P(A,E_1,\tau\rm_{int}(E_{\rm c.m.},J,B)).
\end{eqnarray}
Here, the interaction time $\tau\rm_{int}(E_{c.m.},J,B)$ was obtained from the deflection function method \cite{PhysRevC.27.590}. 
We calculated the fusion probability as
\begin{eqnarray}
P_{\rm CN}(E_{\rm c.m.},J)=\int f(B)P_{\rm CN}(E_{\rm c.m.},J,B)dB
\end{eqnarray}
The Coulomb barrier distribution function $f(B)$ were taken as Eq. \ref{gbd}, so the fusion cross section was written as
\begin{eqnarray}
\sigma\rm _{fus}(E_{\rm c.m.})=\sigma _{\rm cap}(E_{\rm c.m.}) P\rm_{CN}(E\rm_{c.m.},J)  
\end{eqnarray}

\subsection{ Survival probability}

The compound nuclei formed by the all nucleons transfer from projectile nuclei to target nuclei which has a few of excitation energies. The excited compound nuclei was extreme unstable which would de-excited by evaporating $\gamma$-rays, neutrons, protons, $\alpha$ $etc.$) against fission. The survival probability of the channels x-th neutron, y-th proton and z-alpha was expressed as \cite{Chen_2016,FENG201082,FENG200933}
\begin{eqnarray}
&&W\rm_{sur}(E_{CN}^*,x,y,z,J)=P(E\rm_{CN}^*,x,y,z,J)\nonumber\\&& \times \prod\rm_{i=1}^{x}\frac{\Gamma \rm_n(E\rm_i^*,J)}{\Gamma\rm _{tot}(E_i^*,J)} \prod\rm_{j=1}^{y}\frac{\Gamma\rm _p(E_j^*,J)}{\Gamma \rm_{tot}(E\rm_i^*,J)} \prod\rm_{k=1}^{z}\frac{\Gamma\rm _\alpha (E_k^*,J)}{\Gamma \rm _{tot}(E_k^*,J)}  
\end{eqnarray}
where the $E_{\rm CN}^*$ and $J$ were the excitation energy and the spin of the excited nucleus, respectively. The total width $\Gamma_{\rm tot}$ was the sum of partial widths of particles evaporation, $\gamma$-rays and fission. The excitation energy $E_S^*$ before evaporating the $s$-th particles was evaluated by
\begin{eqnarray}
E\rm_{s+1}^*=E\rm_s^* - B _i ^n - B _j ^p - B_k ^\alpha - 2T_s
\end{eqnarray}
with the initial condition $E\rm_i^*$=$E\rm_{\rm CN}^*$ and $s$=$i$+$j$+$k$. The $B\rm_{\rm i}^n$, $B_{\rm j}^p$, $B\rm_{\rm k}^\alpha$ are the separation energy of the $i$-th neutron, $j$-th proton, $k$-th alpha, respectively. The nuclear temperature $T_i$ was defined by $E_{\rm i}^*=\alpha T_{\rm i}^2-T_{\rm i}$ with the level density parameter $a$. The decay width of the $\gamma$-rays and the particle decay were evaluated with the similar method in Ref. \cite{Chen_2016}. We set $E^*-B_{\rm v}-\delta -\delta _{\rm n}$ to the term $\varrho $.

The widths of particles decay were evaluated with the Weisskopf evaporation theory as
\begin{eqnarray}
&&\Gamma_{\rm v}(E^*,J)=(2s_{\rm v}+1)\frac{m_{\rm v}}{\pi ^2\hbar ^2\rho (E^*,J)}\nonumber\\&& \times \int_{0}^{\varrho -\frac{1}{a} }\varepsilon \rho (\varrho +\delta-E_{\rm rot}-\varepsilon ,J) \sigma _{\rm inv}(\varepsilon )d\varepsilon .  
\end{eqnarray}
Here, $s_{\rm v}$, $m_{\rm v}$ and $B_{\rm v}$ are the spin, mass and binding energy of the particle, respectively. The pairing correction energy $\delta$ was set to be $12/\sqrt{A}$, 0, $-12/\sqrt{A}$ for even-even, even-odd and odd-odd nuclei, respectively. The inverse cross section was taken by $\sigma_{\rm inv}=\pi R_{\rm \nu}^{2}T(\nu) $. The penetration probability was set 1 for neutrons and $T(\nu) =(1 + \exp(\pi(V_{\rm C}(\nu)-\varepsilon)/\hbar\omega))^{-1}$ for charged particles with the $\hbar \omega= 5 $ and 8 MeV for proton and $\alpha$, respectively. The Coulomb barrier of the emitting charge particles and daughter nuclei was calculated by
\begin{eqnarray}
V_{\rm C} = \frac{ (Z_{\rm CN}-i) Z_i e^2}{r_i(A_{{\rm CN}-i}^{1/3}+A_i^{1/3})} 
\end{eqnarray}
In this work, we set proton emitting $r_p$ = 1.7 fm, and $\alpha$ emitting $\alpha$ = 1.75 fm, more detail information could be seen in Ref. \cite{PhysRevC.68.014616}.
The fission width was calculated by Bohr-Wheeler formula as in Ref.\cite{PhysRevC.80.057601,PhysRevC.76.044606}. We set $ E^*-B_{\rm f}-E_{\rm rot}-\delta -\delta _{\rm f}$ to the term $ \kappa $.
\begin{eqnarray}
&&\Gamma_f(E^*,J)=\frac{1}{2\pi \rho_f (E^*,J)}\int_{0}^{\kappa -\frac{1}{\alpha _f} }\nonumber \\&&  
\frac{\rho _f(\kappa -\varepsilon+\delta ,J)d\varepsilon }{1+\rm exp\left [ -2\pi (\kappa -\varepsilon+\delta+\delta_f )/\hbar \omega  \right ] } 
\end{eqnarray}
For heavy fragments, the fission width was usually taken as $\hbar\omega=$ 2.2 MeV \cite{artza05}, and $\delta_f$ was the pairing correction for the fission barrier. The fission barrier was divided the microscopic part and the macroscopic part which was written as the following form
\begin{eqnarray}
B\rm_f(E^*,J)=B_f^{LD}+B_f^M(E^*=0,J)\rm exp(-E^*/E_D)
\end{eqnarray}
where the macroscopic part was derived from liquid-drop model, as follows
\begin{eqnarray}
B^{LD}\rm_f = \left \{\begin{array} {rl} 0.38(0.75 - x )E\rm_{s0} & , (1/3 < x < 2/3) \\ \\
0.83(1-x)^3 E\rm_{s0} & ,(2/3 < x < 1) \end{array} \right.
\end{eqnarray}
with 
\begin{eqnarray}
x=\frac{E\rm_{C0}}{2E\rm_{S0}}. 
\end{eqnarray}
Here, $E_{\rm c0}$ and $E_{\rm s0}$ were the surface energy and Coulomb energy of the spherical nuclear, respectively, which could be taken from the Myers-Swiatecki formula
\begin{eqnarray}
E\rm_{s0}=17.944[1-1.7826(\frac{N-Z}{A})^2]A^{2/3} \ MeV 
\end{eqnarray}
and
\begin{eqnarray}
E\rm_{c0}=0.7053\frac{Z^2}{A^{1/3}} \ MeV. 
\end{eqnarray}
Microcosmic shell correction energy were taken from \cite{Moller_1995}. Shell damping energy was
\begin{eqnarray}
E\rm_D=\frac{5.48A^{1/3}}{1+1.3A^{-1/3}} \ MeV  
\end{eqnarray}
or 
\begin{eqnarray}
E\rm_D=0.4A^{4/3}/a \ MeV
\end{eqnarray}
Here $a$ is the energy level density parameter. the fission level density was set as $a_f$ = 1.1$a$. The moments of inertia of fission compound nuclei at the ground state (gs) and the saddle point (sd) configuration are given by
\begin{eqnarray}
\zeta \rm_{gs(sd)}=k\times \frac{2}{5}mr^2(1+\beta _2^{gs(sd)}/3). 
\end{eqnarray}
Here, $k$ = 0.4 and $\beta_2$ was quadrupole deformation which were taken from Ref. \cite{Moller_1995}. The $\beta_2^{sd}$ = $\beta_2^{gs}$ + 0.2 was the quadrupole deformation at the saddle point which was calculated by relativistic mean field theory. Based on the Fermi gas model, the energy level density could be expressed as \cite{Moller_1995}
\begin{eqnarray}
\rho (E^*,J)=K\rm_{coll}\times \frac{2J+1}{24\sqrt{2}\sigma ^3a^{1/4}(E^*-\delta )^{5/4} } \nonumber \\ \times  \exp\left [ 2\sqrt{a(E^*-\delta )}-\frac{(J+1/2)^2}{2\sigma ^2}   \right ]  
\end{eqnarray}
with $\sigma^2 = 6\bar{m}^2\sqrt{a(E^*-\delta)}/\pi^2$ and $\bar{m}\approx0.24A^{2/3}$. The $K_{coll}$ was the collective enhancement factor which contain the rotational and vibration effects. The level density parameter was combined with the shell correction energy $E_{sh}(Z,N)$ and the excitation energy $E^{\ast}$ as
\begin{eqnarray}\label{pld}
&&a(E^{\ast},Z,N) = \tilde{a}(A)[1+E\rm_{ \rm sh}(Z,N)f(E^{\ast})/(E^{\ast})]
\end{eqnarray}
Here, $\tilde{a}(A)=\alpha A + \beta A^{2/3}b_{ \rm s}$ was the asymptotic Fermi-gas value of the level density parameter at high excitation energy. The shell damping factor was given by
\begin{eqnarray}
f(E^{\ast})=1-\exp(-\gamma E^{\ast})
\end{eqnarray}
with $\gamma=\tilde{a}/(\epsilon A^{4/3})$. The $\alpha$, $\beta$, $b_{s}$ and $\epsilon$ were set as 0.114, 0.098, 1. and 0.4, respectively. 

The realization probability of evaporation channels was the important component in the survival probability equation. The realization probability of one particle evaporation was given by 
\begin{eqnarray}
P(E\rm_{CN}^*,J)=\rm exp\left ( -\frac{(E_{CN}^*-B_s-2T)^2}{2\sigma ^2}  \right ) 
\end{eqnarray}
where $\sigma$ was the half-height width of the excitation function of the residual nucleon in the fusion-evaporation reactions which was taken as 2.5 MeV in our calculation. For the multiple neutrons evaporation channels $(x>1)$, the realization probability could be derived by the Jackson formula, as follows
\begin{eqnarray}
P(E_{\rm CN}^*,s,J)=I(\bigtriangleup _s,2s-3)-I(\bigtriangleup _{s+1},2s-1)
\end{eqnarray}
where the quantities $I$ and $\bigtriangleup $ were given by 
\begin{eqnarray}
I(z,m)=\frac{1}{m!}\int_{0}^{z}u^me^{-u}du  
\end{eqnarray}
\begin{eqnarray}
\bigtriangleup _s=\frac{E_{CN}^*- {\textstyle \sum_{i=1}^{s}B_i^v} }{T_i} 
\end{eqnarray}
The $B_i^v$ is the separation energy of the evaporation of the $i$-th particle and $s$=$x+y+z$. The spectrum of realization probabilities determines the distribution shape of survival probability in the evaporation channels.

\section{Results and Discussion}\label{sec3}

\begin{figure*}[htb]
\includegraphics[width=1.\linewidth]{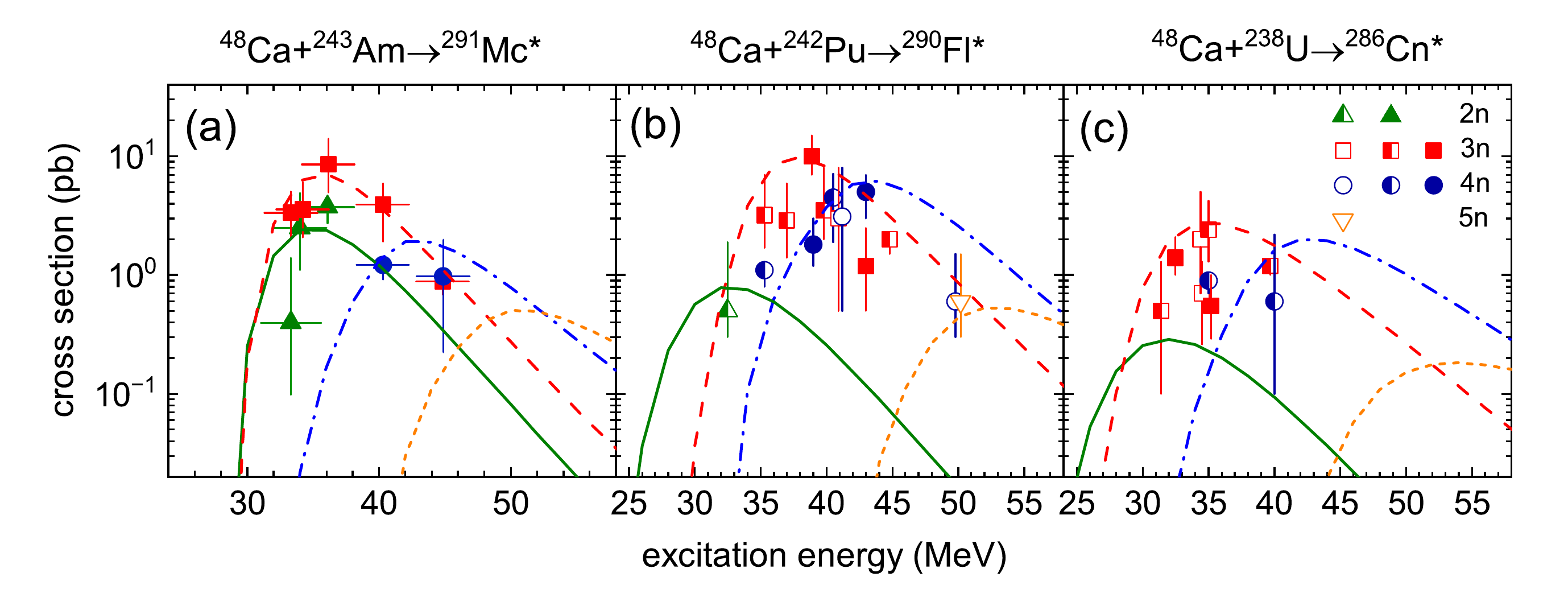}
\caption{\label{fig1} (Color online)
The calculated excitation functions of 2n-, 3n-, 4n- and 5n-evaporation channels for the reactions of $^{48}$Ca+$^{243}$Am, $^{48}$Ca+$^{242}$Pu and $^{48}$Ca+$^{238}$U are marked by solid olive, dash red, dash-dot blue and orange short-dash lines, respectively.
The experimental measure result of excitation functions for 2n-, 3n-, 4n- and 5n-evaporation channels are represented by up-triangle, square, circle and down-triangle. Vertical error bars correspond to total uncertainties. Symbols with arrows show upper cross-section limits. Data marked by open, half-closed, and filled symbols are taken from \cite{PhysRevC.106.024612,PhysRevC.70.064609,hofarticle,JPSJ.86.085001,PhysRevLett.105.182701,PhysRevC.87.014302,PhysRevC.69.021601,PhysRevLett.108.022502,PhysRevC.72.034611}, respectively. 
}
\end{figure*}

In the framework of the DNS model involving all of the collision orientations, we have calculated the excitation functions of 2n-, 3n-, 4n-, 5n-evaporation channels for the collisions of $^{48}$Ca+$^{243}$Am, $^{48}$Ca+$^{242}$Pu and $^{48}$Ca+$^{238}$U marked by solid olive, dash red, dash-dot blue and orange short-dash lines, respectively, as shown in Fig. \ref{fig1}.
In the penal (a), the olive-filled up-triangle, red-filled square, and blue-filled circle stand for the experiment results of 2n-, 3n-, 4n-evaporation channels for the $^{48}$Ca+$^{243}$Am taken from Ref. \cite{PhysRevC.69.021601,PhysRevC.87.014302,PhysRevLett.108.022502}. From reference \cite{PhysRevC.69.021601}, the experiments of $^{48}$Ca+$^{243}$Am at incident energies $E_{\rm lab} = 248, 253$ MeV were carried out at FLNR, JINR. At $E_{\rm lab} = 248$ MeV, three similar decay chains consisting of five consecutive $\alpha$ decays.
At $E_{\rm lab} = 253$ MeV, the decay properties of these synthesized nuclei are consistent with consecutive $\alpha$ decays originating from the parent isotopes of the new element Mc, $^{287}$Mc and $^{288}$Mc, produced in the 3n- and 4n-evaporation channels with cross sections of about 3 pb and 1 pb, respectively.
From the reference \cite{PhysRevC.87.014302}, the cross section for the 3n-evaporation channel reaches its maximum $\sigma_{\rm 3n}$ = $8.5^{+6.4} _{-3.7}$ pb at $E^*$ = 34.0 – 38.3 MeV and decreases with further increase of the excitation energy of the compound nucleus $^{291}$Mc. At the excitation energy, $E^*$ = 44.8 $\pm$ 2.3 MeV, not a single event indicating the formation of $^{288}$Mc was detected. The upper cross-section limit can thus be set at the level $\sigma_{\rm 3n}$ $\leq$ 1 pb. 
At excitation energy in the range of $E^*$ = 31.1 - 36.4 MeV, the cross sections for the formation of ERs in the 3n- and 2n-evaporation channels were about $3.5^{+2.7}_{-1.5}$ pb and $2.5^{+2.7}_{-1.5}$ pb, respectively. 
At energies $E^*$ $\leq$ 36 MeV, as it could be expected for the 2n-evaporation product, $^{289}$Mc was not detected. The upper cross-section limit can be set at the level $\sigma_{\rm 2n}$ $\leq$ 3 pb.
From the reference \cite{PhysRevLett.108.022502}, the cross sections for the formation of ERs in the 3n- and 2n-evaporation channels are about $3.2^{+0.8}_{-1.2}$ pb and $0.3^{+0.7}_{-0.2}$ pb at energies $E^*$ = 33 MeV, respectively.
In Fig. \ref{fig1} (b), the olive up-triangle, red square, blue circle, and orange down-triangle stand for the experiment results of 2n-, 3n-, 4n-, 5n-evaporation channels for the $^{48}$Ca+$^{242}$Pu, respectively, where filled, half-filled, and open symbols represent three experiments for the $^{48}$Ca+$^{242}$Pu \cite{PhysRevLett.105.182701,PhysRevC.106.024612,PhysRevC.70.064609,PhysRevLett.103.132502}. From the Ref. \cite{PhysRevC.106.024612}, a maximum cross section of $10.4 ^{+3.5}_{-2.1}$ pb was measured for the $^{242}$Pu($^{48}$Ca, 3n)$^{287}$Fl reaction. From the Ref. \cite{PhysRevLett.105.182701}, at excitation energy $E^*$ = 50 MeV, the $^{242}$Pu($^{48}$Ca, 5n)$^{285}$Fl cross section was $0.6 ^{+0.9}_{-0.5}$ pb. The no-observation of a 3n evaporation product gave an upper limit for the $^{242}$Pu($^{48}$Ca, 3n)$^{285}$Fl reaction of 1.1 pb. The 3n and 4n cross section values measured at $E^*$ = 41 MeV were $3.1 ^{+4.9}_{-2.6}$ pb.
In Fig. \ref{fig1} (c), the red square, blue circle stand for the experiment results of 3n-, 4n-evaporation channels for the $^{48}$Ca+$^{238}$U, respectively, where filled, half-filled, and open symbols represent three experiments for the $^{48}$Ca+$^{238}$U \cite{hofarticle,PhysRevC.106.024612,PhysRevC.70.064609,JPSJ.86.085001}. 
From the Ref. \cite{PhysRevC.70.064609}, the maximum cross section values of the xn-evaporation channels for the reaction $^{238}$U($^{48}$Ca, xn)$^{286-x}$Cn were measured to be $\sigma_{\rm 3n}$ = $2.5^{+1.8}_{-1.1}$ pb and $\sigma_{\rm 4n}$ = $0.6^{+1.6}_{-0.5}$ pb. At the excitation energies of the compound nucleus $E^*$ = 34.5 MeV, two decay events from $^{283}$Cn were observed, resulting in the cross section of $2.0^{+2.7}_{-1.3}$ pb\cite{JPSJ.86.085001}. The cross-section deduced from all four events is $0.72^{+0.58}_{-0.35}$ pb, measured at the excitation energy of 34.6 MeV of the compound nucleus $^{286}$Cn\cite{hofarticle}.
From the above three panels, we can see that our calculations have a good agreement with the available experimental excitation functions of the reactions $^{48}$Ca+$^{243}$Am, $^{48}$Ca+$^{242}$Pu and $^{48}$Ca+$^{238}$U.
\begin{figure*}[htb]
\includegraphics[width=1.\linewidth]{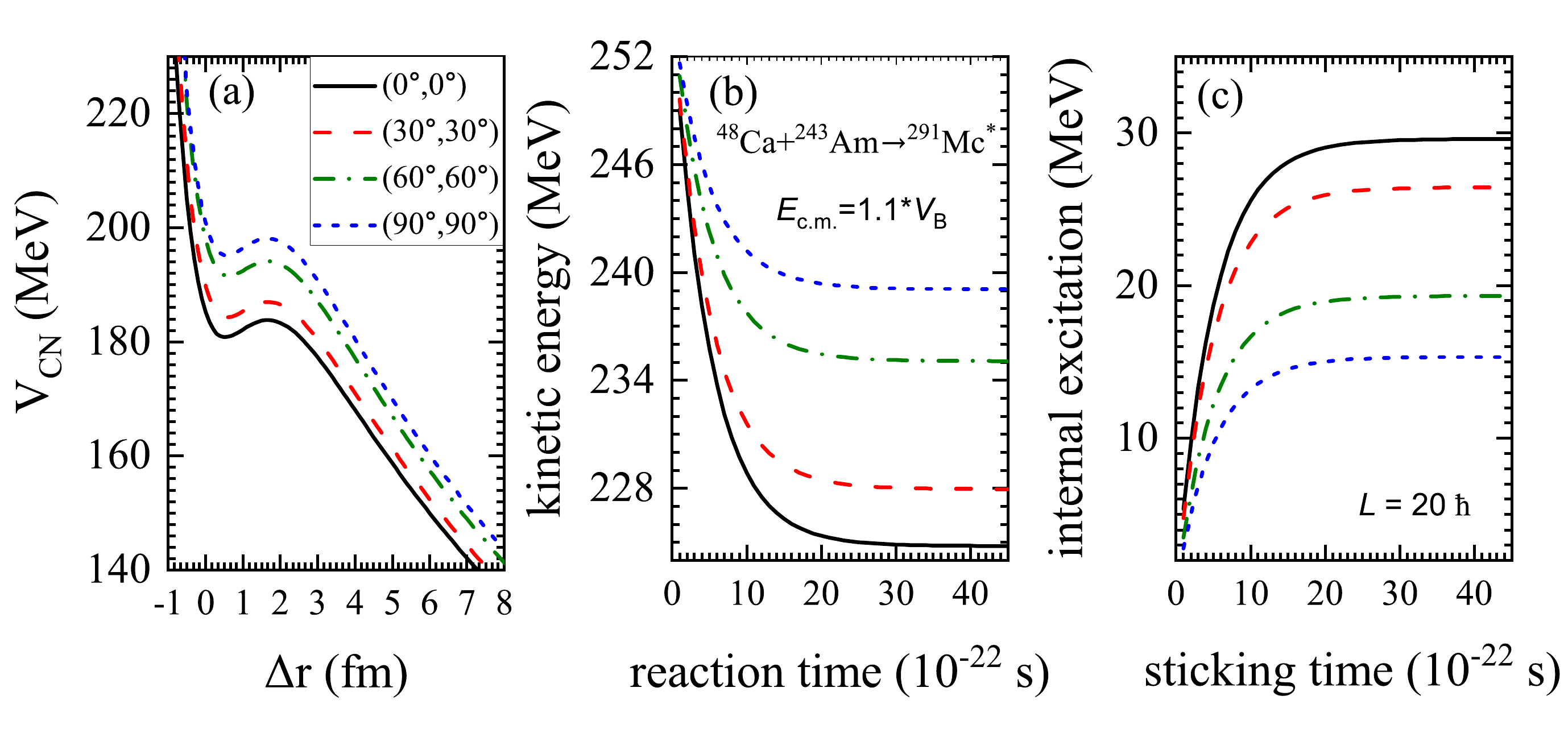}
\caption{\label{fig2} (Color online) Panel (a) shows the interaction potential for the collisions of $^{48}$Ca+$^{243}$Am as a function of distance with different collision angles. The collision orientation $(0^{\circ}, 0^{\circ})$, $(30^{\circ}, 30^{\circ})$, $(60^{\circ}, 60^{\circ})$ and $(90^{\circ}, 90^{\circ})$ correspond to solid black, red dash, olive dash-dot and short dash lines, respectively. Panel (b) represents the time-varying function of radial kinetic energy dissipating in the collision system under the angular momentum $L=20$ $\hbar$. Panel (c) exhibits the internal excitation energy of the composite system varies with the sticking time with the given angular momentum $L=20$ $\hbar$.
}
\end{figure*}

To investigate the dependence of production cross section of superheavy nuclei in the fusion-evaporation reactions on collision orientation, we have exported four configurations of the collision orientations from our calculations for the reaction of $^{48}$Ca+$^{243}$Am as $(0^{\circ}, 0^{\circ})$, $(30^{\circ}, 30^{\circ})$, $(60^{\circ}, 60^{\circ})$ and $(90^{\circ}, 90^{\circ})$ marked by the solid black, red dash, olive dash-dot and blue short-dash lines, respectively in Fig. \ref{fig2}. The projectile nuclei $^{48}$Ca and target nuclei $^{243}$Am have theoretical quadrupole deformation value $\beta_{\rm P} = 0.$ and $\beta_{\rm T} = 0.224$. In Fig. \ref{fig2}, panel (a) shows the distributions of interaction potential energy to the distance between the surfaces of projectile nuclei and target nuclei. The interaction potential $V_{\rm CN}$ consists of Coulomb potential $V_{\rm C}$ and nucleus-nucleus potential $V_{\rm N}$, which were calculated by Wong formula \cite{PhysRevLett.31.766} and double folding method \cite{PhysRevC.69.024610}, respectively. The interaction potential energies are increased with the large collision orientations, because of the large effective interaction face. The panel (b) displays the distributions of relative radial kinetic energy to interaction time. The kinetic energy decreased exponentially with the increased reaction time, for the given impact parameter $L=20$ $\hbar$.
The evolution reach equilibrium at $2\times 10^{-21}$s. The equilibrium kinetic energy are 225 MeV, 228 MeV, 235MeV and 239 MeV, corresponding to collision orientations $(0^{\circ}, 0^{\circ})$, $(30^{\circ}, 30^{\circ})$, $(60^{\circ}, 60^{\circ})$ and $(90^{\circ}, 90^{\circ})$, respectively. The kinetic energy were dissipating into internal excitation of the composite system, correspondingly which were increased exponentially with the reaction time and have the same relaxation time, as illustrated in panel (c). 
From Fig. \ref{fig2}, we can see that the interaction potential and the evolution of kinetic energy and internal excitation energy were highly dependent on the orientations, actually, which were the basic reasons lead to the dependence of final synthesis cross sections of superheavy nuclei on collision orientations.

\begin{figure*}[htb]
\includegraphics[width=1.\linewidth]{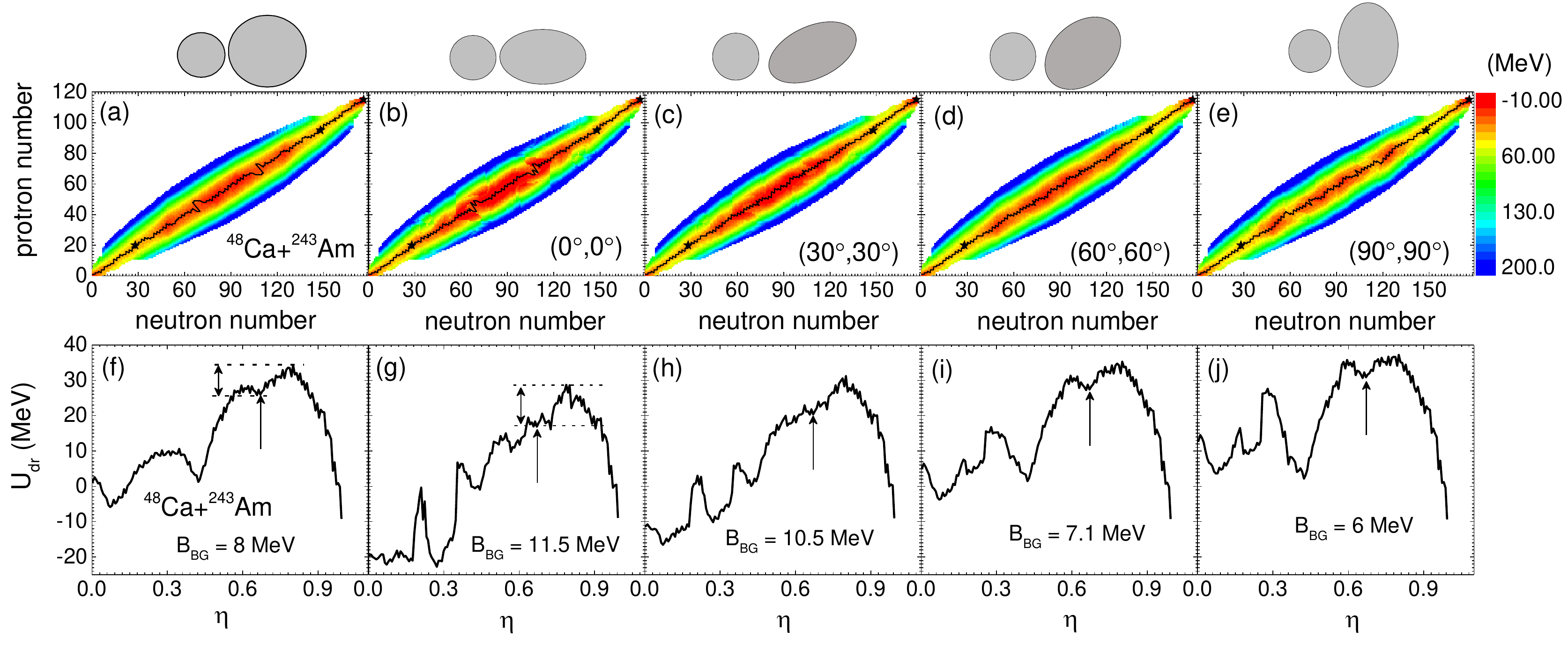}
\caption{\label{fig3}(Color online) The panels (a), (b), (c), (d), and (e) represent the potential energy surface (PES) of $^{48}$Ca+$^{243}$Am at collision orientations of sphere-sphere, $(0^{\circ}, 0^{\circ})$, $(30^{\circ}, 30^{\circ})$, $(60^{\circ}, 60^{\circ})$ and $(90^{\circ}, 90^{\circ})$, respectively. The panels (f), (g), (h), (i), and (j) correspond to their collision orientation-based valley trajectories in PES, respectively. Their inner barrier value are shown with $B_{\rm BG}$ The arrow lines stand for the injection points.
}
\end{figure*}

The potential energy surface (PES) and driving potential (DP) of the reaction $^{48}$Ca+$^{243}$Am were calculated by Eq. \ref{pes} for the collision orientations of sphere-sphere, $(0^{\circ}, 0^{\circ})$, $(30^{\circ}, 30^{\circ})$, $(60^{\circ}, 60^{\circ})$ and $(90^{\circ}, 90^{\circ})$, as illustrated in Fig. \ref{fig3}. 
The PES and DP were listed in the upper layer panels and lower panels, respectively. The panel (a) and (f) shown the PES and DP of the no-deformation of the projectile-target nuclei. The minimum trajectories and injection points were attached to the PES, which were represented by solid black lines and filled black stars. The panels (a) and (f) were the PES and DP of no-deformation collision. The structure effect were clearly shown in the PESs and DPs, by the comparison of no-deformation collision and with duadrupole deformation collision. The inner fusion barrier were taken as the difference between the injection points and  Businaro-Gallone (B.G.) points, which were 8 MeV, 11.5 MeV, 10.5 MeV, 7.1 MeV and 6 MeV corresponding to collision orientations of no-deformation, $(0^{\circ}, 0^{\circ})$, $(30^{\circ}, 30^{\circ})$, $(60^{\circ}, 60^{\circ})$ and $(90^{\circ}, 90^{\circ})$, respectively. It was found that the inner fusion barrier were highly dependent on the collision orientations, which could reveal the fusion probability directly. The inner fusion barriers were decrease with the increase collision orientation which has the smallest value at the waist-waist collision. The sketches of collision orientations were illustrated in the top of Fig. \ref{fig3}. The potential energy of the symmetry field in the PES increase along with the increasing collision orientations, because the corresponding Coulomb force increases.  

\begin{figure*}[htb]
\includegraphics[width=1.\linewidth]{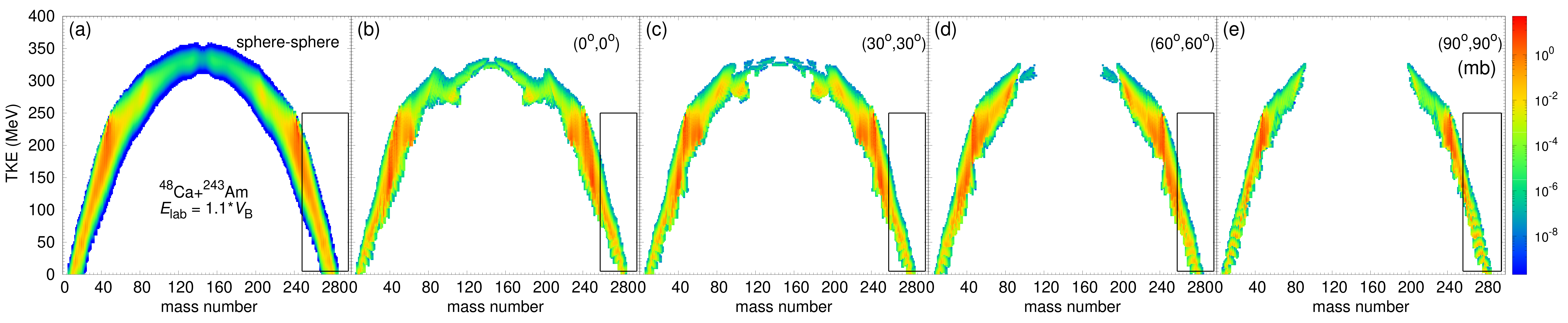}
\caption{\label{fig4} (Color online) The panels (a), (b), (c), (d) and (e) display
the calculations of TKE-mass distribution of the primary fragments in the collisions of $^{48}$Ca+$^{248}$Am at $E_{\rm c.m.}$ = 1.1$\times V_{\rm B}$ for their collision orientations of sphere-sphere, $(0^{\circ}, 0^{\circ})$, $(30^{\circ}, 30^{\circ})$, $(60^{\circ}, 60^{\circ})$ and $(90^{\circ}, 90^{\circ})$, respectively.}
\end{figure*}

In collision process, overcome the Coulomb barrier, the kinetic energies of the colliding partners dissipate into the composite system rapidly. The probability of projectile and target diffuse along the PES, which were calculated by solving a set of master equations. The total kinetic energy (TKE) of binary fragments were related to the incident energy, ground state binding energy, internal excitation energy as $TKE = E_{\rm cm} - V_{\rm cn} - Q_{\rm gg} - E^*$. The Fig. \ref{fig4} presents the TKE-mass distributions for collision orientations of no-deformation, $(0^{\circ}, 0^{\circ})$, $(30^{\circ}, 30^{\circ})$, $(60^{\circ}, 60^{\circ})$ and $(90^{\circ}, 90^{\circ})$ at incident energy $E_{\rm lab}$ = $1.1\times V_{\rm B}$, as shown in panels (a), (b), (c), (d) and (e), respectively. The TKE term could be rewritten as $TKE = E_{\rm cm} - U_{\rm dr} - E^*$. So The TKE-mass distribution shape were highly dependent on the driving potential. The TKE-mass distribution for no-deformation collision in panel (a) were more smooth than others in panels (b), (c), (d) and (e) which shown the structure effect in the TKE-mass distribution. The fragments in the black square passed the B.G. points were supposed to lead to fusion. The fusion probability were calculated by summing all of the formation probability passed through B.G. point. From the Fig. \ref{fig4}, it was hard to value the dependence of fusion probability on the collision orientations, because it only show the one incident energy $E_{\rm lab}$ = $1.1\times V_{\rm B}$. 

\begin{figure*}[htb]
\includegraphics[width=.9\linewidth]{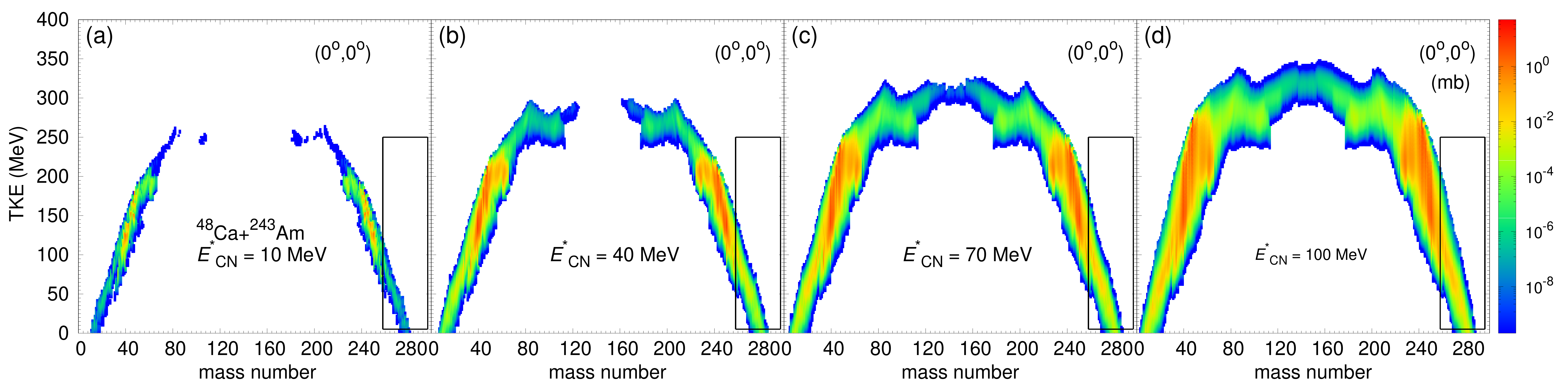}
\caption{\label{fig5} (Color online)
The panels (a), (b), (c) and (d) show the calculations of TKE-mass distribution of the primary fragments in the head-on collisions of $^{48}$Ca+$^{248}$Am at their incident energies correspond to excitation energies of compound nuclei 10 MeV, 40 MeV, 70 MeV and 100 MeV, respectively.
}
\end{figure*}

Figure \ref{fig5} shows the TKE-mass distributions at the excitation energies $E^*_{CN}$ = 10 MeV, 40 MeV, 70 MeV, 100 MeV for the tip-tip collisions of $^{48}$Ca+$^{248}$Am, as illustrated in the panels (a), (b), (c) and (d), respectively. From Fig. \ref{fig5}, we can see that the TKE-mass distribution were broader along the increased incident energy. Obviously, the fusion probability were increased along the larger excitation energy. However, the compound nuclei with large excitation energy could lead to fission easily. The maximum evaporation residue cross section of the highly excitation compound nuclei were the balance between the fusion probability and the survival probability.

\begin{figure*}
\includegraphics[width=.95\linewidth]{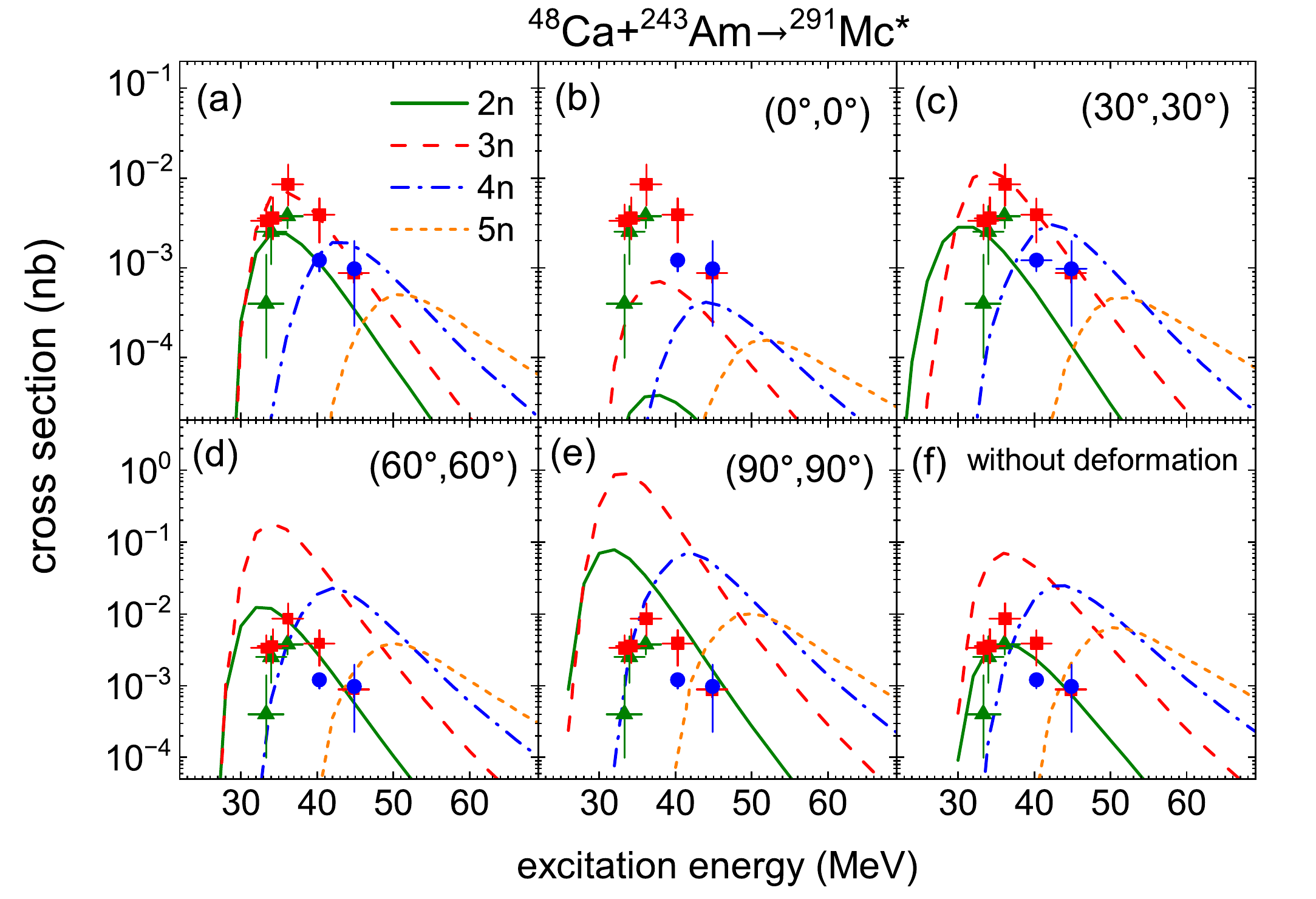}
\caption{\label{fig6} (Color online) In the collisions of $^{48}$Ca+$^{248}$Am, the panels show the calculations of excitation functions in 2n-, 3n-, 4n- and 5n- evaporation channels correspond to solid olive, red dash, blue dash-dot and orange short-dash lines, respectively. 
Panels (b), (c), (d), (e) and (f) display the excitation functions at orientations $(0^{\circ}, 0^{\circ})$, $(30^{\circ}, 30^{\circ})$, $(60^{\circ}, 60^{\circ})$, $(90^{\circ}, 90^{\circ})$ and sphere to sphere, respectively.
Panel (a) show the total excitation function in taking all of the collision orientations by the method of Gaussian distribution.
Experimental data are marked by filled up-triangle, square, circle and down-triangle symbols are taken from \cite{PhysRevC.87.014302,PhysRevC.69.021601}.
}
\end{figure*}
As far as possible to approach the real collision process, we proposed a Gaussian-like barriers distributions used to consider all of the collision orientations, which has the same formula in Eq. (\ref{gbd}).
The olive solid lines, red dash lines, blue dot-dash lines and orange short-dash lines stand for the calculated excitation function of the 2n-, 3n-, 4n-, 5n-evaporation channels. The olive filled up-triangle, red filled square, blue filled circle represent the experimental excitation function of the 2n-, 3n-, 4n-evaporation channels, respectively.
For the reactions $^{48}$Ca+$^{248}$Am at the excitation energy interval of $E^*$ = 1-100 MeV, the excitation function of the 2n-, 3n-, 4n-, 5n-evaporation channels were calculated by the DNS model involving the barrier distribution, as shown in panel (a), which were in a nice agreement with the experiment data \cite{PhysRevC.87.014302,PhysRevC.69.021601}. The calculated excitation functions of $^{48}$Ca+$^{248}$Am for the collision orientations $(0^{\circ}, 0^{\circ})$, $(30^{\circ}, 30^{\circ})$, $(60^{\circ}, 60^{\circ})$, $(90^{\circ}, 90^{\circ})$ and no-deformation were listed in panels (b), (c), (d), (e), (f), respectively. It was found that the $(0^{\circ}, 0^{\circ})$ collisions underestimate the experiment results. The collisions $(30^{\circ}, 30^{\circ})$ fit the experiment results relatively nice. The collisions $(30^{\circ}, 30^{\circ})$, $(60^{\circ}, 60^{\circ})$ and $(90^{\circ}, 90^{\circ})$ overestimate the experiment data. From the Fig. \ref{fig6}, It was found that the DNS model involving barrier distributions could reproduce the experimental results quite well.

\begin{figure*}
\includegraphics[width=1.\linewidth]{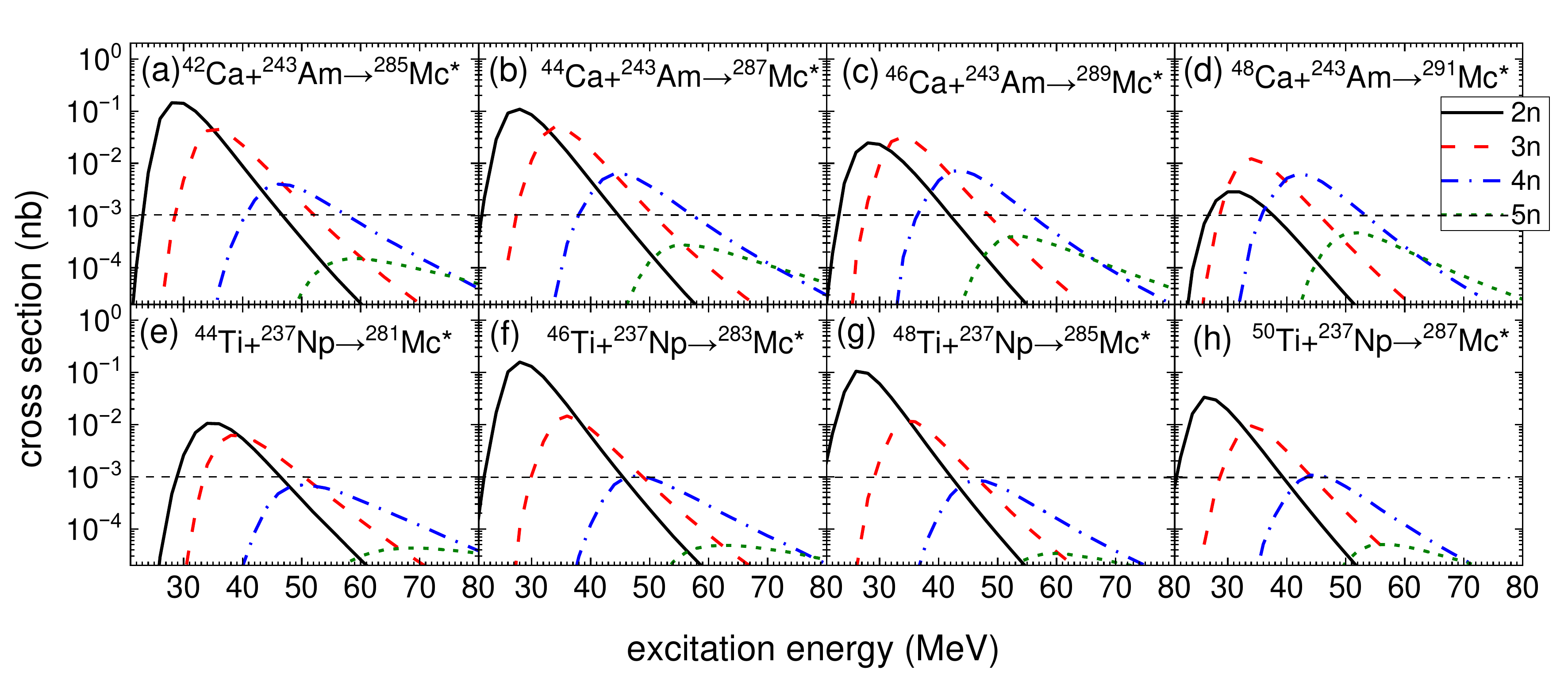}
\caption{\label{fig7} (Color online)
The calculations of excitation functions in the collisions of $^{42}$Ca+$^{243}$Am, $^{44}$Ca+$^{243}$Am, $^{46}$Ca+$^{243}$Am, $^{48}$Ca+$^{243}$Am, $^{44}$Ti+$^{237}$Np, $^{46}$Ti+$^{237}$Np, $^{48}$Ti+$^{237}$Np and $^{50}$Ti+$^{237}$Np shown in panels (a), (b), (c), (d), (e), (f), (g), and (h), respectively.
The 2n-, 3n-, 4n- and 5n- evaporation channels correspond to black solid, red dash, blue dash-dot and olive short-dash lines, respectively
}
\end{figure*}
Based on the DNS model involving barrier distribution, to investigate the dependence of evaporation residue cross section on the isospin of projectile, we have calculated the reactions of $^{42}$Ca+$^{243}$Am, $^{44}$Ca+$^{243}$Am, $^{46}$Ca+$^{243}$Am, $^{48}$Ca+$^{243}$Am, $^{44}$Ti+$^{237}$Np, $^{46}$Ti+$^{237}$Np, $^{48}$Ti+$^{237}$Np and $^{50}$Ti+$^{237}$Np at the excitation energies interval of $E^* = 1 - 80 $ MeV systematically. 
From the Fig. \ref{fig7}, it was found that the excitation functions of evaporation residue cross section were highly dependent on the isospin of projectile. For the isotopes of Ca induced reactions, the cross sections of 2n-, 3n-evaporation channels were decrease along with the projectile of Ca isotopes with large N/Z, which might cause by fusion probability. The ratio of $\sigma_{\rm 3n}/\sigma_{\rm 2n}$ were increase along with the increasing N/Z, which shown that more-neutron-rich compound nuclei prefer to evaporate more neutrons. The existed moscovium isotopes were $^{287-290}$Mc. The predictions of maximum cross sections of the new $^{281-286}$Mc were 4 pb, 45 pb, 150 pb, 50 pb, 101 pb and 30 pb, respectively in the calcium isotopes induced fusion-evaporation. The maximum synthesis cross section of new moscovium isotopes was $^{283}$Mc as 0.15 nb in the reactions $^{42}$Ca+$^{243}$Am. For the Ti isotopes induced reactions, the 2n-evaporation channel were dominant in the evaporation residue cross sections. The maximum synthesis cross section of Mc was $^{281}$Mc as 0.2 nb in the reactions $^{46}$Ti+$^{237}$Np. The new moscovium isotopes of $^{278-286}$Mc were evaluated as 0.5 pb, 9 pb, 12 pb, 10.5 pb, 150 pb, 11 pb, 100 pb, 10 pb, 31 pb, respectively, in the titanium isotopes induced fusion-evaporation.

\begin{figure*}
\includegraphics[width=1.\linewidth]{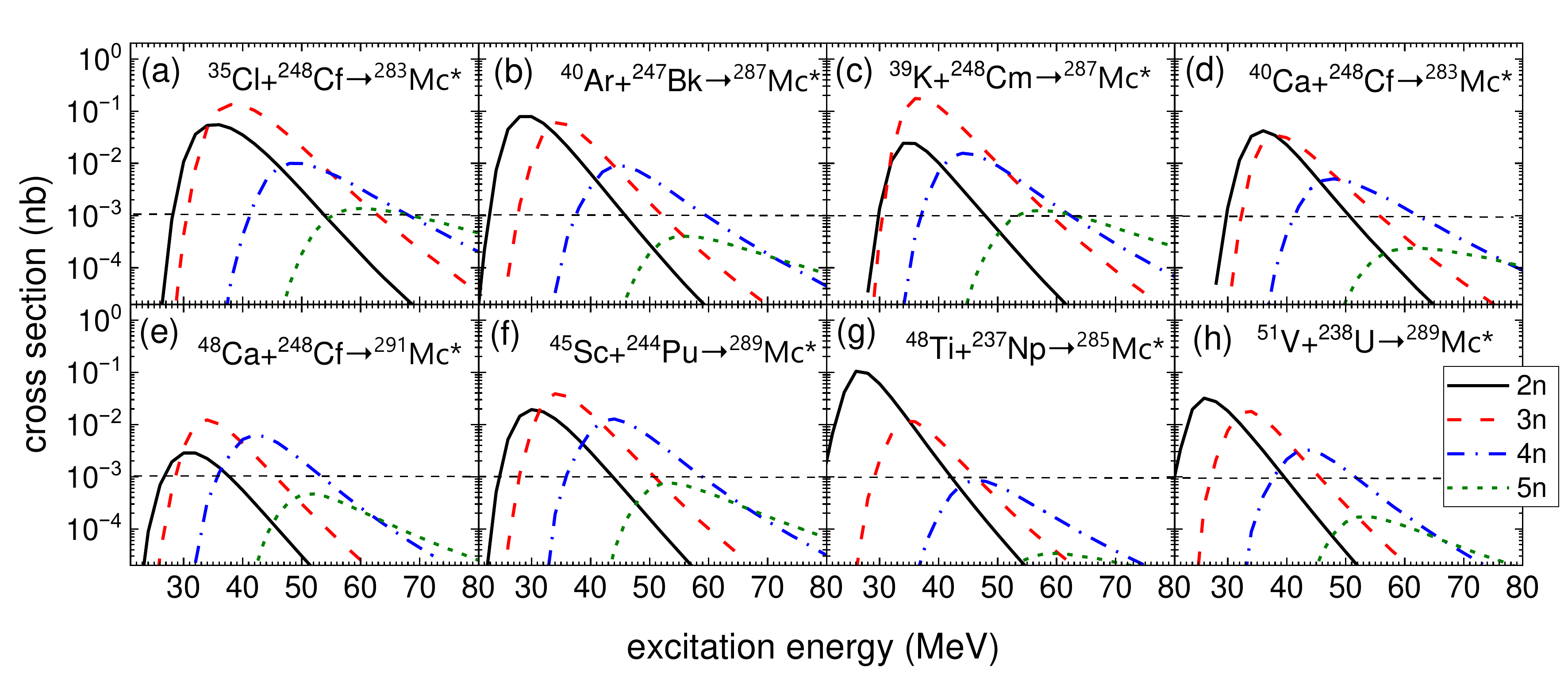}
\caption{\label{fig8} (Color online)
The calculations of excitation functions in the collisions of $^{35}$Cl+$^{248}$Cf, $^{40}$Ar+$^{247}$Bk, $^{39}$K+$^{247}$Cm, $^{40}$Ca+$^{243}$Am, $^{48}$Ca+$^{243}$Am, $^{45}$Sc+$^{244}$Pu, $^{48}$Ti+$^{237}$Np and $^{51}$V+$^{238}$U shown in panels (a), (b), (c), (d), (e), (f), (g), and (h), respectively.
The 2n-, 3n-, 4n- and 5n- evaporation channels correspond to black solid, red dash, blue dash-dot and olive short-dash lines, respectively.
}
\end{figure*}
To investigate the influence of entrance effect on the synthesis cross section of superheavy moscovium in the fusion-evaporation reactions, we have calculated the collisions of the $^{35}$Cl + $^{248}$Cf ($\eta$ = 0.75), $^{40}$Ar + $^{247}$Bk ($\eta$ = 0.72), $^{39}$K + $^{247}$Cm ($\eta$ = 0.73), $^{40}$Ca + $^{243}$Am ($\eta$ = 0.72), $^{48}$Ca + $^{243}$Am ($\eta$ = 0.67), $^{45}$Sc + $^{244}$Pu ($\eta$ = 0.69), $^{48}$Ti + $^{237}$Np ($\eta$ = 0.66) and $^{51}$V + $^{238}$U ($\eta$ = 0.65) systematically based on the DNS model, as illustrated in panels (a), (b), (c), (d), (e), (f), (g) and (h), respectively. The mass asymmetry $\eta$ were respect to $\eta = (A_{\rm T} - A_{\rm P}) / (A_{\rm T} + A_{\rm P})$. From the Fig. \ref{fig8}, we can see that the reaction systems with large $\eta$ prefer to produce large production cross section because the large mass asymmetry reactions were favor to fusion. In these calculations, the new moscovium $^{278-286}$Mc have been predicted as the production cross section value 1 pb, 10 pb, 130 pb, 50 pb, 15 pb, 100 pb, 30 pb, 200 pb, 40 pb, respectively.
The 2n- or 3n-evaporation residue channels were dominant in the evaporation survival process. The ratio of $\sigma_{\rm 3n}/\sigma_{\rm 2n}$ illustrated the role of odd-even effect on the production cross section of superheavy nuclei. 
The maximum production cross section of moscovium isotopes was predicted as 200 pb in the reaction $^{247}$Cm($^{39}$K, 3n)$^{283}$Mc. 

\section{Conclusions}\label{sec4}

Summarizing, as far as possible to simulate the real collision process, we have proposed a Gaussian-like barrier distribution function used to including all collision orientations. To investigate the dependence of production cross section of superheavy isotopes on the collision orientations, we have calculated the reactions of $^{48}$Ca+$^{243}$Am at the excitation energies interval of 0-100 MeV for the collision orientations no-deformation, $(0^{\circ}, 0^{\circ})$, $(30^{\circ}, 30^{\circ})$, $(60^{\circ}, 60^{\circ})$ and $(90^{\circ}, 90^{\circ})$ systematically. In the DNS model, for given collision orientation, some physical quantities such as interaction potential, radial kinetic energy, internal excitation energy, TKE-mass, potential energy surface, driving potential and inner fusion barrier were exported to show the influence of collision orientations which were highly dependent on the collision orientations. We have compared the calculated excitation functions of $^{48}$Ca+$^{243}$Am at some fixed collisions orientations and the available experimental results. We found that the large collision orientations shown the overestimated value, compared to experiment data. The collision orientation nearby $(30^{\circ}, 30^{\circ})$ could fit the experiment data very well. The barrier distribution-based excitation function were in a good agreement with the experiment data. To test the barrier distribution function, we have calculated the reactions of $^{48}$Ca+$^{243}$Pu and $^{48}$Ca+$^{238}$U, which have nicely reproduced the experimental excitation functions. 
Based on the DNS model involving the barrier distribution function, we have calculated the reactions of projectiles $^{42-48}$Ca bombarding on target $^{243}$Am and projectiles $^{42-48}$Ca on target $^{237}$Np systematically. The influence of the isospin of projectile on production cross section have been studied. For Ca induced F.E. reactions, the $\sigma_{\rm 2n}$ and $\sigma_{\rm 3n}$ were dominant in the evaporation residue cross sections which were decrease along with the increase N/Z in projectiles. The ratio of the $\sigma_{\rm 3n}/\sigma_{\rm 2n}$ increase along with the increase N/Z in projectiles, which might be caused by neutron-rich compound nuclei were favor to lose neutrons. For Ti induced F.E. reactions, the maximum cross section were 150 pb own by $^{283}$Mc in the reaction $^{237}$Np($^{46}$Ti, 2n)$^{283}$Mc.
The reactions of $^{35}$Cl+$^{248}$Cf, $^{40}$Ar+$^{247}$Bk, $^{39}$K+$^{247}$Cm, $^{40}$Ca+$^{243}$Am, $^{48}$Ca+$^{243}$Am, $^{45}$Sc+$^{244}$Pu, $^{48}$Ti+$^{237}$Np and $^{51}$V+$^{238}$U have been calculated to investigate the entrance channel effect on production cross sections of superheavy nuclei. The large mass asymmetry systems lead to the large production cross section. we also found that odd-even effect might paly a role in the evaporation residue cross section. We have predicted the new moscovium isotopes $^{278-286}$Mc with the maximum cross sections 0.5 pb, 9 pb, 12 pb, 10.5 pb, 150 pb, 11 pb, 100 pb, 10 pb, 31 pb in the collisions of $^{35,37}$Cl + $^{248}$Cf, $^{38,40}$Ar + $^{238}$Bk, $^{39,41}$K + $^{247}$Cm, $^{40,42,44,46}$Ca + $^{238}$Am, $^{45}$Sc + $^{242}$Pu, and $^{46,48,50}$Ti + $^{243}$Np, $^{51}$V + $^{238}$U at excitation energy interval of 0-100 MeV.

%

\end{document}